\documentclass[twocolumn,preprintnumbers,amsmath,amssymb]{revtex4}
\usepackage{graphicx}
\usepackage{dcolumn}
\usepackage{bm}
\usepackage{hyperref} 
\usepackage{multirow} 
\usepackage{soul}
\newcommand\beq{\begin{equation}}
\newcommand\eeq{\end{equation}}
\newcommand\bea{\begin{eqnarray}}
\newcommand\eea{\end{eqnarray}}
\usepackage{wrapfig}
\newcommand{\nonum}{\nonumber}

\begin{document}

\title{\bf Bond order wave (BOW) phase of the extended Hubbard model: 
Electronic solitons, paramagnetism, coupling to Peierls and Holstein phonons\\}
\author{\bf Manoranjan Kumar and Zolt\'an G. Soos }
\address{ Department of Chemistry, Princeton University, Princeton NJ 08544 \\}
\date{\today}

\begin{abstract}
The bond order wave (BOW) phase of the extended Hubbard model (EHM) 
in one dimension (1D) is characterized at intermediate correlation $U = 4t$ 
by exact treatment of $N$-site systems. Linear coupling to lattice 
(Peierls) phonons and molecular (Holstein) vibrations are treated in 
the adiabatic approximation. The molar magnetic susceptibility $\chi_M(T)$ 
is obtained directly up to $N = 10$. The goal is to find the consequences 
of a doubly degenerate ground state (gs) and finite magnetic gap $E_m$ in a 
regular array. Degenerate gs with broken inversion symmetry are 
constructed for finite $N$ for a range of $V$ near the charge density wave 
(CDW) boundary at $V \approx 2.18t$ where $E_m \approx 0.5t$ is large. The 
electronic amplitude $B(V)$ of the BOW in the regular array is shown to mimic 
a tight-binding band with small effective dimerization $\delta_{eff}$. Electronic 
spin and charge solitons are elementary excitations of the BOW phase and also 
resemble topological solitons with small $\delta_{eff}$. Strong infrared 
intensity of coupled molecular vibrations in dimerized 1D systems is shown 
to extend to the regular BOW phase, while its temperature dependence is related 
to spin solitons. The Peierls instability to dimerization has novel aspects for 
degenerate gs and substantial $E_m$ that suppresses thermal excitations. 
Finite $E_m$ implies exponentially small $\chi_M(T)$ at low temperature followed by 
an almost linear increase with $T$. The EHM with $U = 4t$ is representative of 
intermediate correlations in quasi-1D systems such as conjugated polymers or 
organic ion-radical and charge-transfer salts. The vibronic and thermal properties 
of correlated models with BOW phases are needed to identify possible physical realizations. 
\vskip .4 true cm
\noindent PACS numbers: 71.10.Fd, 73.22.Gk, 75.40.Cx, 78.30.-j \\
\noindent Email: soos@princeton.edu
\end{abstract}

\maketitle

\section{Introduction}
Nakamura identified the bond order wave (BOW) phase \cite{r1} 
of the one-dimensional (1D) half-filled extended Hubbard model 
\cite{r2} (EHM, Eq. \ref{eq1}). The key features are broken 
inversion symmetry, doubly degenerate ground state (gs) and 
finite magnetic gap $E_m$ in a regular (equally spaced) array. 
Competition among electron delocalization $t$, on-site repulsion 
$U > 0$ and nearest-neighbor repulsion $V > 0$ stabilizes the BOW 
phase over a narrow range whose boundaries motivated subsequent 
studies \cite{r3,r4,r5,r6,r7,r8,r9}. The BOW phase has 
$V \approx U/2$ and substantial $t$. The quantum transition to the 
charge density wave (CDW) phase at large $V$ is first order \cite{r4,r5} for 
$U > U^* \approx 7t$, continuous for $U < U^*$. The BOW/CDW 
boundary is at $V_c(U)$ for $U < U^*$, while the boundary to the 
spin-fluid phase with $E_m = 0$ is at $V_s(U) < V_c(U)$. Other 
half-filled 1D Hubbard models with spin-independent interactions 
have narrow BOW phases between the spin-fluid and CDW phases \cite{r10}.

Theoretical studies of the EHM have focused on the quantum phase diagram 
of an extended array without reference to possible physical realizations.
 Even for models, however, 1D instabilities and finite temperature must 
be addressed. The principal goal of this paper is to characterize the 
BOW phase of the EHM at $U = 4t$, a typical choice for intermediate 
correlation, with special attention to the consequences of gs degeneracy 
and finite $E_m$. Coupling to Peierls or Holstein phonons requires gs 
derivatives in the adiabatic (Born-Oppenheimer) approximation, and finite 
temperature properties are needed to assess physical realizations. It is 
clearly desirable to understand BOW phases prior to specific applications. 
Broadly similar properties are expected for other $U$ or other 1D models with 
spin-independent interactions. 
 
Quasi-1D organic molecular crystals and conjugated polymers have strong 
electron-phonon (e-ph) coupling and electron-molecular-vibration (e-mv) 
coupling in addition to intermediate correlation. The Su-Schrieffer-Heeger 
(SSH) model of polyacetylene has linear e-ph coupling and topological 
solitons as elementary excitations \cite{r11,r12}. The Peierls instability 
is driven by e-ph coupling and has spectacular e-mv consequences in infrared 
spectra when inversion symmetry is broken. Conjugated polymers and organic 
ion-radical salts have been a playground for 1D Hubbard models 
\cite{r12,r13,r14,r15,r16,r17} with e-ph and e-mv coupling, variable 
electron or hole filling, degenerate or nondegenerate gs, and either 
segregated or mixed stacks. The bandwidth is $4t \approx 10$ eV in 
polymers and $4t \approx 1$ eV in $\pi$-stacks, with comparable 
intermediate $t/U$. The Peierls instability of Hubbard-type models is a 
rich separate topic \cite{r18}. We are not aware of work on 
either e-ph or e-mv coupling in the BOW phase. 
As shown below, finite $E_m$ and degenerate gs lead to electronic solitons 
in a rigid regular array that nevertheless resemble SSH solitons. With suitable 
modification, extensive SSH analysis \cite{r17,r12} can be applied to models with a BOW phase.

We recently proposed that a BOW phase is realized in Rb-TCNQ(II), 
the second polymorph of a tetracyanoquinodimethane salt \cite{r19,r20}. 
The evidence is a 100 K crystal structure ({\it P\={1}}) with regular 
{$ \rm TCNQ^- $ } stacks at inversion centers, negligible spin 
susceptibility below 140 K that indicates a large $E_m$, 
and infrared spectra that demonstrates broken electronic inversion symmetry. 
Broken $C_i$ symmetry and finite $E_m$ in a regular 1D array are precisely 
the signatures of a BOW phase \cite{r1}. Large $E_m$ indicates proximity 
to the CDW boundary, and we will so choose $V$ in the EHM at $U = 4t$. 
We have Rb-TCNQ(II) and alkali-TCNQs in mind, but do not model them 
explicitly beyond invoking solitons for the temperature dependence of 
infrared spectra. In addition to values of microscopic parameters, 
interactions between chains must be addressed in actual models along with 
the Coulomb interactions and transfer integrals for different stacking 
motifs. We consider electronic properties of the EHM at $U = 4t$ with 
linear coupling to Peierls and Holstein phonons. Vibrational degrees of 
freedom are introduced as needed in the adiabatic
approximation. \\

The EHM describes electronic degrees of freedom in a regular 1D 
array \cite{r2}

\begin{eqnarray}
H_{el} & =& \sum^N_{p=1,\sigma} -t(  a^{\dagger}_{p,\sigma}  a_{p+1,\sigma} + h.c) \nonum \\
       & + & \sum^N_{p=1} ( U n_p(  n_p-1)/2 +V  n_p n_{p+1}) \label{eq1}
\end{eqnarray}

\noindent where $t = 1$ is the unit of energy, h.c. is the hermitian conjugate, 
$a^{\dagger}_{p,\sigma} (a_{p\sigma})$ creates (annihilates) an 
electron with spin $ \sigma$ at site $p$, and $n_p$ is the 
number operator. $H$ conserves total spin. The gs is a singlet 
$(S = 0)$ for $U$, $V \ge 0$ and even $N$. The half-filled band 
with $N$ electrons and $N$ sites has electron-hole (e-h) symmetry 
$J = \pm 1$ and inversion symmetry $C_i$ at sites that we label as 
$\sigma = \pm 1$. The correlated many-electron basis increases 
as $\approx 4^N $ at large $N$. Valence bond (VB) methods \cite{r21,r22} 
at present yield exact results up to $N = 17$ for low-energy states 
and the full spectrum up to $N = 10$. Nakamura identified \cite{r1} 
the BOW phase using field theory, symmetry arguments and numerical 
results up to $N = 12$.\\

\begin{figure}[h]
\begin {center}
\hspace*{-0cm}{\includegraphics[width=8.5cm,height=9.5cm,angle=-90]{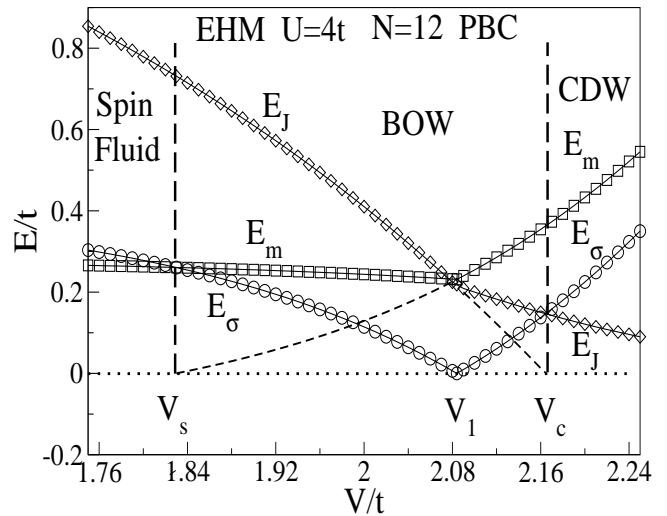}} \\
\caption{ Excitation thresholds and crossovers of the 12-site extended Hubbard model, 
Eq. \ref{eq1}, with periodic boundary conditions. $E_m$, $E_{\sigma}$ and $E_J$ 
are energy gaps to the lowest triplet and the lowest singlets with opposite $C_i$ 
and e-h symmetry, respectively. $V_s$ is defined by $E_{\sigma} = E_m$, $V_1$ by $E_{\sigma} = 0$ 
and $V_c$ by $E_{\sigma} = E_J$. }
\label{fig1}
\end {center}
\end{figure}
To introduce the EHM phase diagram at 0 K, we define three threshold 
excitations from the singlet gs: the magnetic gap $E_m$ to the lowest triplet, 
the gap $E_J$ to the lowest singlet with opposite $J$, and the gap $E_\sigma$ 
to the lowest singlet with opposite $C_i$. Increasing $V$ at constant $U$, $t$ 
drives the system from a spin-fluid phase to a CDW phase. Fig. \ref{fig1} 
shows the evolution of $E_m$, $E_J$ and $E_{\sigma}$ with $V$ for $U = 4t$, 
$N = 12$. The BOW phase spans $V_s < V < V_c$, where $V_s(N)$ and $V_c(N)$ 
are defined by the excited-state crossovers $E_{\sigma}(N) = E_m(N)$ and 
$E_{\sigma}(N) = E_J(N)$, respectively. Table \ref{tb1} lists the remarkably weak 
size dependence of $V_s(N)$ and $V_c(N)$ for $N = 4n$ with periodic boundary 
conditions (PBC, $t_{1N} = 1$) and $N = 4n+2$ with antiperiodic boundary 
conditions ($t_{1N} = -1$). Similarly, the $V_s$ boundary of a 
frustrated spin chain has been found \cite{r23} using $E_m(N) = E_{\sigma}(N)$ up 
to $N = 24$. Multiple methods yield the boundary $V_c(N)$ of the CDW phase \cite{r7,r10}. 
The points $V = V_1(N)$ in Table \ref{tb1} or Fig. \ref{fig1} correspond to $E_{\sigma} = 0$. 
They mark a gs degeneracy that is central to our discussion, where broken 
symmetry gs are readily constructed.\\

\begin{table}
\caption { Excitation thresholds and crossovers of the extended Hubbard model, Eq. \ref{eq1}, 
with $N$ sites, $U = 4$, and $t_{1N} = \pm 1$ for $N = 4n$, $4n + 2$.}
\begin{tabular}{cccc} \hline
$N$ ~~~& $V_s ~~~(E_{\sigma} = E_m)~~~$ & $V_1(E_{\sigma} = 0)~~~$ & $ V_c (E_{\sigma} = E_J)$  \\\hline
8 ~~~ &1.8094 ~~~&2.0597~ ~~&2.1592 \\
   &       &        &      \\
10~~~ & ~~~~~~1.8190 & ~~~~~~~~~2.0726 & ~~~~~~~~~2.1624 \\
   &       &        &      \\
12 ~~~& 1.8297 ~~& 2.0840 ~~~& 2.1645\\
   &       &        &      \\
14~~~ & ~~~~~~1.8311 & ~~~~~~~~~2.0925 & ~~~~~~~~~2.1651\\
   &       &        &      \\
16~~~ & 1.8452~~ & 2.0981 ~~~& 2.1653 ~~~\\\hline

\end{tabular}
\label{tb1}
\end{table}

The paper is organized as follows. Broken inversion symmetry and elementary 
excitations are treated exactly in Section II for finite $N$. 
Electronic solitons, both spin and charge, are found in regular chains 
with open boundary conditions (OBC, $t_{1N} = 0$) and compared to SSH solitons. 
Section III deals with linear coupling to molecular (Holstein) and 
lattice (Peierls) vibrations. The Berry phase formulation of polarization 
is applied to the infrared activity of molecular vibrations when $C_i$ 
symmetry is broken. The Peierls instability of the BOW phase is contrasted to 
the SSH model within the limitations of an adiabatic approximation. The magnetic 
gap $E_m$ and spin susceptibility $\chi_M$ of the BOW phase are obtained in 
Section IV for large $E_m$ close to the CDW instability. 
Large $E_m$ reduces the thermal population of excited states and opens a new 
regime in which spin solitons govern $\chi_M$. The discussion in Section V 
summarizes the consequences of broken symmetry and finite $E_m$ such as the 
temperature dependence of the infrared intensity or of $ \chi_M$. We 
briefly mention extensions to BOW phases of related models.

\section{Broken symmetry and elementary excitations }

We discuss the EHM, Eq. \ref{eq1} with $U = 4$, $t = 1$, 
using exact results for finite $N$ with PBC($t_{1N} = 1$) for $N = 4n$ and $t_{1N} = -1$ 
for $N = 4n + 2$. The gs kinetic energy is given by the bond orders $p_n$ of successive sites

\begin{eqnarray}
2p_n  =\langle \psi_0| \sum_{\sigma} ( a^{\dagger}_{n,\sigma}  a_{n+1,\sigma} + h.c)| \psi_0 \rangle.
 \label{eq2}
\end{eqnarray}

\noindent Broken $C_i$ symmetry in the BOW phase leads to $p_{2n} \ne p_{2n-1}$ , while broken e-h symmetry 
in the CDW phase leads to different electron count $n_{2p} \ne n_{2p-1}$ 
in the even and odd sublattice. At constant $U$ and $t$, the order parameter $B(V)$ of the BOW phase is

\begin{eqnarray}
B(V)=|p_{2n}(V)-p_{2n-1}(V)|.
 \label{eq3}
\end{eqnarray}
$B(V)$ is large between sites $2n$, $2n - 1$ in one broken-symmetry gs and between $2n$, $2n + 1$ in the other.\\

Since both $E_m$ and $E_{\sigma}$ vanish rigorously in the spin-fluid phase with $V < V_s$, 
finite gaps in Fig. \ref{fig1} are due to finite $N$. The dashed line 
$E_m(V) - E_{\sigma}(V)$ is an approximation for opening the magnetic gap. 
Similarly, the dashed line $E_J(V) - E_{\sigma}(V)$ approximates the closing of the e-h gap at $V_c$. 
Finite $N$ limits $E_{\sigma} = 0$ to points $V_1(N)$ in Fig. \ref{fig1} and Table \ref{tb1}. 
At gs crossovers, we construct broken-symmetry states
 
\begin{eqnarray}
 |\psi_{\pm} (V_1) \rangle=(|\psi_{\sigma=1}\rangle \pm |\psi_{\sigma=-1}\rangle )/\sqrt 2
 \label{eq4}
 \end{eqnarray}
and compute their bond orders in Eq. \ref{eq2}. Degenerate gs at finite $N$ 
provide  direct access to BOW systems that is not available for Monte Carlo, which yields the gs energy, 
or density matrix renormalization group (DMRG) calculations. DMRG with OBC breaks $C_i$ 
symmetry for even $N$ and returns a nondegenerate gs. We are not aware of DMRG with PBC that 
conserves $C_i$ and gives the energy and gs in both the $\sigma= \pm 1$ sectors. It is easier to 
find an energy crossover at $V_1(N)$ where $E_{\sigma}= 0$ for finite $N$ than to 
demonstrate a degenerate gs over the interval $V_s < V < V_c$ in the extended system.\\

On the other hand, we need the gs degeneracy at $U = 4t$ beyond the single point $V_1(N)$. 
To do so, we add the following perturbation to Eq. \ref{eq1}

\begin{eqnarray}
H(J_2)= J_2 \sum_p \vec S_p. \vec{S}_{p+2}.
 \label{eq5}
\end{eqnarray}

\begin{figure}[h]
\begin {center}
\hspace*{-0cm}{\includegraphics[width=8.5cm,height=9.5cm,angle=-90]{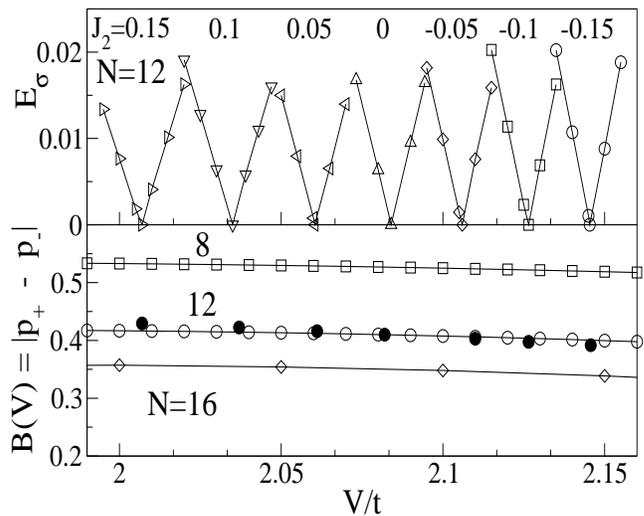}} \\
\caption{  (top panel) Energy $E_{\sigma}$ of the lowest singlet with opposite $C_i$ 
symmetry in a 12-site EHM with $U=4t$ and PBC as a function of $V$ in Eq . \ref{eq1} 
and the indicated $J_2$ values in Eq. \ref{eq5}. (bottom panel) Order parameter $B(V_1,J_2) = |p_+ - p_-|$ 
in Eq. \ref{eq3} at $E_{\sigma }= 0$ for $N = 12$ (closed symbols) and $B(V)$ 
for $N$ = 8, 12, 16 for $J_2$ = 0 (open symbols).  }
\label{fig2}
\end {center}
\end{figure}

\noindent $H(J_2)$ acts on second neighbors with one electron each. 
The upper panel of Fig. \ref{fig2} shows how $V_1(J_2)$ scans $E_{\sigma}=0$ 
across the BOW phase at $U = 4t$, lowering $V_1$ for antiferromagnetic $J_2 > 0$ 
and raising it for $J_2 < 0$. The interval $J_2 = \pm 0.15$ in Fig. \ref{fig2} 
is sufficient to enforce $E_{\sigma}(J_2) = 0$ between $V/t = 2.0$ and 
2.15 for the EHM with $N = 12$ and $U/t = 4$. Strictly degenerate gs in the $\sigma= \pm 1$ 
sectors will be essential for electron vibrational coupling in Section III.\\

The gs in Eq. \ref{eq4} and the order parameter $B$ in Eq. \ref{eq3} 
hold for $V_1(J_2)$ where $E_\sigma= 0$. The lower panel of Fig. \ref{fig2} compares $B(V_1(J_2))$ 
at the indicated $J_2$ values for $N = 12$ with $B(V)$ based on $J_2 = 0$, when $| \psi_{\pm} \rangle$ 
in Eq. \ref{eq4} are linear combinations of functions that are not quite degenerate. 
$B(V,N)$ depends weakly on $E_{\sigma}$ and $N$ in this interval. Of course, $H(J_2)$ 
also perturbs other properties, at least slightly, and becomes a strong perturbation 
for $V < 2t$. We are interested in $V$ close to $V_c$, where $E_\sigma$ is 
small and $E_m$ is large. \\

Bond orders $p_{\pm}(V,N)$ and $B(V,N)$ are well approximated near 
$V \approx V_1(N)$ without invoking   $J_2$, and results in this Section are based on $J_2 = 0$. 
For example, $U = 4t$, $V = 2.05t$, $N = 16$ returns $p_+ = 0.752$, $p_- = 0.398$, $B = 0.354$. As expected \cite{r24} near 
the metallic point $V_c$, the average $p = 0.574$ is within $10\%$ of $2/\pi$, 
the value for a tight binding (H\"uckel) band of free electrons with $U = V = 0$ 
in Eq. \ref{eq1}.
The $t$ term dominates near $V_c$ where $U$ and $V$ 
almost cancel.\\ 

The SSH model \cite{r11,r12} is a tight-binding band with linear 
e-ph coupling $ \alpha = (dt/du)_0$ to the Peierls phonon, the optical 
mode of the 1D chain, and an adiabatic approximation for a harmonic 
lattice. Its gs is dimerized, with 
$t_n = -(1 -\delta (-1)^n)$ along the stack. Bond orders of the 
infinite chain are readily found analytically, with $p_+(\delta) = 0.806$, 
$p_-(\delta) = 0.452$ and $B(\delta) = p_+ - p_- = 0.354$ at $\delta = 0.10$ 
that matches $B(V)$ of the finite EHM above. The origin of broken $C_i$ symmetry 
is quite different: e-ph coupling in SSH, correlation $U$,$V$ in EHM. 
Moreover, $B(V)$ opens at $V_s \approx 1.86t$ and is almost constant for $V > 2.0t$ 
before vanishing abruptly at $V_c \approx 2.18t$ while $B(\delta)$ is monotonic in $\delta$. 
The electronic gs are nevertheless similar and previous discussions of 
topological solitons or domain walls between regions with opposite $B$ can be applied to BOW 
phases \cite{r12,r17,r15,r25}.\\ 
\begin{figure}[h]
\begin {center}
\hspace*{-0cm}{\includegraphics[width=7.5cm,height=9.5cm,angle=-90]{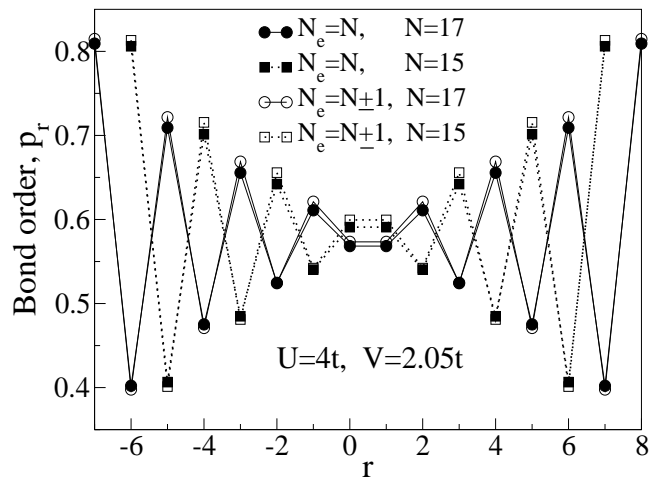}} \\
\caption{ Bond orders $p_r$ in Eq. \ref{eq2} of a spin soliton (close symbols) or 
charge soliton (open symbols) in an $N$-site EHM with $t_{1N} = 0$, $U = 4t$ and $V = 2.05t$ in Eq.\ref{eq1}. }
\label{fig3}
\end {center}
\end{figure}

We start with electronic domain walls and return later to the Peierls instability. 
Domain walls are modeled as in SSH with odd $N$ in Eq. \ref{eq1} and $t_{1N} = 0$. We retain PBC for $V$ 
to minimize end effects. $N_e = N$ 
electrons correspond to a neutral soliton with $S = 1/2$, while $N \pm 1$ electrons 
are charge solitons with $S = 0$. Spin-charge relations are reversed just as 
in SSH. We label sites from $r = 0$ at the center to $r = \pm(N-1)/2$ 
at the ends. Positive and negative solitons have equal bond orders by e-h symmetry. 
We could solve $N = 17$ exactly using a symmetry-adapted valence bond basis to compute $p_n$ 
in Eq. \ref{eq2}. As shown in Fig. \ref{fig3} for $N = 15$ and 17, the 
bond orders of spin and charge solitons are symmetric about the center. 
They are slightly larger at the chain ends than $p_- = 0.398$, $p_+ = 0.752$, 
the values of the BOW phase at $V = 2.05t$. This end effect for $t_{1N} = 0$ 
also appears for even $N$ and alternating bond orders with largest $p_+$ at either end.\\

As context for these results, we note that 
Eq. \ref{eq1} with $U = V = 0$ and odd $N$ can be read as a H\"{u}ckel model of an alternant 
hydrocarbon with a nonbonding orbital at $\epsilon = 0$ that is empty in the cation, 
singly occupied in the radical and doubly occupied in the anion. Since the $\epsilon = 0$ 
orbital has nodes at every other site for any $N$, it does not contribute to bond 
orders that are consequently equal for neutral and charge solitons in the SSH model. 
The correlated model has equal $p_r$ for charge solitons that differ slightly from 
the spin soliton. In every case, bond orders are $p_+$ at the ends and reverse smoothly in between. \\ 

\begin{figure}[h]
\begin {center}
\hspace*{-0cm}{\includegraphics[width=8.5cm,height=9.5cm,angle=-90]{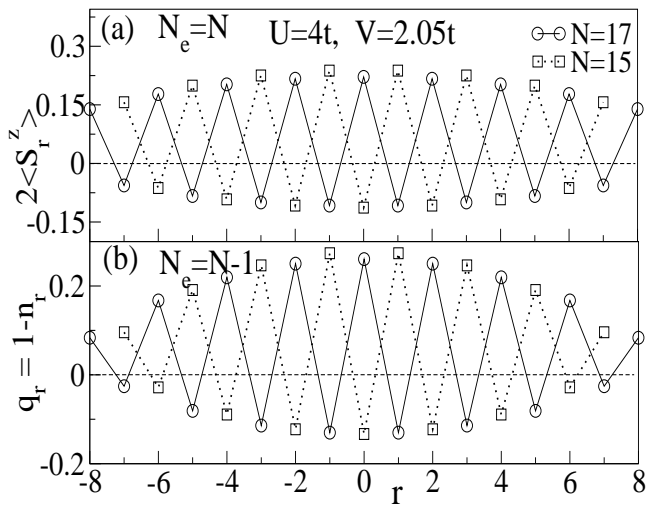}} \\
\caption{(a) Soliton spin density $2\langle S^z_r \rangle$ and (b) 
soliton charge density $q_r = 1 - n_r$ of an $N$-site EHM with $t_{1N} = 0$, $U = 4t$ and $V = 2.05t$ 
in Eq. \ref{eq1}.} 
\label{fig4}
\end {center}
\end{figure}

Next we compute spin densities $\rho_r = 2\langle S^z_r \rangle $ 
for the radical and charge densities $q_r = 1 - n_r$ for ions. 
The upper panel of Fig. \ref{fig4} has $\rho_r $ for $N = 15$ and 
17 while the lower panel has $q_r$ for the cation; the anion charges are $-q_r$ 
by e-h symmetry. The spin or charge density vanishes in alternant 
hydrocarbons where the nonbonding orbital has nodes, at odd $r$ 
for $N = 4n + 1$ and even $r$ for $N = 4n -1$. As seen in Fig. \ref{fig4}, 
correlation \cite{r25} generates small negative $\rho_r$ at these 
sites and also small $q_r$ of opposite sign. The spin or charge density 
is large in the middle and decreases at the ends. Electronic solitons in 
the BOW phase differ in this respect from a regular tight-binding band, which 
has equal $\rho_r$ or $q_r$ at every other site for any odd $N$. To mimic the BOW 
results in Fig. \ref{fig4} with SSH solitons, finite $\delta \approx 0.05$ is 
needed for $\rho_r$ and $\delta \approx 0.08$ for $q_r$ since large $\delta$ gives faster decrease. 
Electronic solitons in Figs. \ref{fig3} and \ref{fig4} connect regions with $\pm B$ in a regular chain.\\ 

Domain walls or topological solitons are the elementary excitations of the BOW phase. They 
resemble SSH solitons in an appropriately dimerized lattice. The BOW phase has different 
energy for creating a pair of spin or charge solitons. We take $E_0(N,N_e)$ 
as the gs energy of Eq. \ref{eq1} for even $N$, $N_e$ electrons  and $t_{1N} = 0$. 
The formation energy of a pair of spin solitons ($S = 1/2$) is

\begin{eqnarray}
2W_S(N)&=& E_0(N+1,N+1)+ E_0(N-1,N-1) \nonum \\
&-&2 E_0(N,N). 
 \label{eq6}
\end{eqnarray}

\noindent In the limit of large $N$, parallel spins are the lowest 
triplet with $2W_S = E_m$, the magnetic gap for PBC. The results for $N =$ 8, 12 and 
16 in Table \ref{tb2} are for $V = V_1(N)$ in Table \ref{tb1}, where $E_{\sigma} = 0$, 
but the $V$ dependence is weak. The formation energy for a pair of charge solitons is
\begin{eqnarray}
2W_C(N)= E_0(N+1,N)+ E_0(N-1,N)-2 E_0(N,N). 
 \label{eq7}
\end{eqnarray}

\noindent The cation, anion and neutral system are singlets with different charge. The natural comparison is to the charge gap, 

\begin{eqnarray}
2E_C(N)= E_0(N,N+1)+ E_0(N,N-1)-2 E_0(N,N). 
 \label{eq8}
\end{eqnarray}

\noindent For large $N$, $E_C(N)$ is the energy of separated ion radicals. 
As seen in Table \ref{tb2}, $2W_C$ approaches $E_C$ from below, 
as expected since charge solitons are singlets while the ions in $E_C$ 
are doublets. On the other hand, $2W_C$ is larger than $E_J$ because $V > 0$ 
favors adjacent sites with $n = 0$ (hole) and $n = 2$ (electron).\\ 

\begin{table}
\begin{center}
\caption { Representative EHM energies in units of $t$ for $U = 4$ and $V = V_1(N)$ in Eq. \ref{eq1}.}  
\begin{tabular}{cccc} \hline
 Energy, $t = 1$ ~~~ &  $N = 16$ ~~~&  $N = 14$ ~~~&  $N = 12$~~~ \\\hline
$2W_S(N)$ $ \rm Eq. \ref{eq6} $ ~~~& 0.2116~~~ & 0.2403~~~ & 0.2790~~~\\
$2W_C(N)$ $ \rm Eq. \ref{eq7} $ ~~~ & 0.9254~~~ & 1.0281~~~ & 1.1691~~~ \\
$2E_C(N)$ $ \rm Eq. \ref{eq8} $ ~~~ & 1.1669~~~ & 1.2931~~~ & 1.4610~~~ \\ 
$E_m$                         ~~~     & 0.2124~~~ & 0.2233~~~ & 0.2368~~~ \\
$E_J$                         ~~~     & 0.3102~~~ & 0.3046~~~ & 0.2971~~~ \\
$E_3$\footnote{Third lowest singlet}~~~& 0.8959~~~   & 1.0094 ~~~& 1.1572~~~ \\\hline
\end{tabular}
\label{tb2}
\end{center}
\end{table}

Valence bond methods yield low-energy excitations 
in every symmetry sector with fixed $S$, $J$ and $\sigma$.  
Except for total wave vector $k = 0$ or $\pi$, degeneracy in $\pm k$ is 
expected and found. Finite-size effects increase with energy and 
it becomes progressively more difficult to extract more than 3-4 
states for large N. The entries in Table \ref{tb2} are from a much larger 
set. An ``effective'' $\delta_{eff} \approx 0.05 - 0.10$ is inferred from the 
order parameter $B(V)$ or from spin or charge solitons in the EHM with $U = 4t$ 
and $2.0 < V/t < V_c$. While BOW-phase results are for a regular stack, 
they are naturally related to SSH results with $\delta < 0.10$ that is considerably smaller than 
$\delta = 0.18$ based on the optical gap of polyacetylene\cite{r11}.

\section{Coupling to Holstein and Peierls phonons}

The SSH model invokes linear e-ph coupling, $\alpha= (dt/du)_0$, to 
characterize the Peierls instability and elementary excitations of a half-filled 
tight-binding band. Linear coupling to molecular vibrations is the basis for interpreting 
polarized infrared spectra of $\pi$-radical stacks. The operators  
$a^{\dagger}_{p \sigma}$, $a_{p,\sigma}$ create, annihilate 
electrons with spin $\sigma$ in the lowest unoccupied molecular orbital of TCNQ, with 
equal energy $\Delta = 0$ and $n_p = 1$ for a ${\rm TCNQ^-}$ stack.  Charge 
fluctuations \cite{r26} modulate $\Delta$ and illustrate linear Holstein coupling $g_n$ 
to the $ \rm n^{th} $ totally symmetric (ts) molecular vibration. When $C_i$ symmetry 
is broken, ts modes become strongly IR allowed by borrowing intensity from 
the optical charge-transfer excitation and they are polarized along the chain. 
Accordingly, the appearance of ts modes in polarized IR yields detailed 
information \cite{r27,r28,r29} about e-mv coupling constants $g_n$ and has been 
widely used to infer dimerization. Charge fluctuations break e-h symmetry, 
and strict degeneracy at $V_1(N)$ is critical because $\Delta$ modulation 
is a small energy. \\

The Berry-phase formulation \cite{r30,r31} of polarization makes possible 
improved vibronic analysis of extended systems. It is directly applicable to quantum 
cell models \cite{r32} such as the EHM. The polarization $P$ per unit charge and unit length is a phase \cite{r31,r32}
\begin{eqnarray}
P_N= \frac{2\pi}{N}Im(lnZ_N) 
 \label{eq9}
\end{eqnarray}

\begin{eqnarray}
Z_N= \langle \psi_0| exp(2\pi i M/N) |\psi_0 \rangle  
 \label{eq10}
\end{eqnarray}

\begin{eqnarray}
M=\sum^N_{p=1} p(n_p-1) 
\label{eq11}
\end{eqnarray}
where $| \psi_0 \rangle$  is the exact 
gs of an $N$-site supercell and $M$ is the conventional dipole operator 
for a regular array with unit spacing. $|Z_N| = 0$ is a metallic point 
at which $P$ is not defined \cite{r30}; it corresponds to $V_c$ for $U < U^*$ 
and a continuous CDW transition \cite{r10}. The EHM has real $Z$ and $P = 0$ by 
either $C_i$ or e-h symmetry. \\ 

The required generalization of Eq. \ref{eq1} for Holstein coupling in the adiabatic 
approximation for molecular sites is \cite{r32}

\begin{eqnarray}
H(\Delta)= H_{el}+\Delta \sum^N_{p=1} (-1)^p n_p. 
\label{eq12}
\end{eqnarray}
Finite $\Delta $ breaks e-h symmetry, but not $C_i$ symmetry. The gs in the BOW phase 
is now $|\psi_{\pm}(V,\Delta) \rangle$  with $J_2$ in Eq. \ref{eq5} 
chosen to have $E_{\sigma} = 0$ for $\Delta = 0$. Strictly degenerate $\sigma = \pm 1$ 
ensures first-order correction in $\Delta$. Finite $\Delta$ mixes in excited states such as $E_J$ 
in Fig. \ref{fig1} whose energy in the BOW phase decreases rapidly with 
increasing $V$. We again vary $V$ at $U =$ 4, $t = 1$. Aside 
from a multiplicative constant, the IR intensity goes as \cite{r33}
\begin{eqnarray}
I_{IR}(V)=\big( \frac{\partial P(V,\Delta)}{ \partial \Delta}\big )^2_0.
\label{eq13}
\end{eqnarray}
Charge fluctuations give a finite derivative at $\Delta = 0$. 
The IR intensity in Eq. \ref{eq13} is purely electronic. 
It can be partitioned among ts modes as discussed \cite{r27,r28,r29} in systems where dimerization breaks $C_i$ symmetry.\\ 
\begin{figure}[h]
\begin {center}
\hspace*{-0cm}{\includegraphics[width=8.5cm,height=9.5cm,angle=-90]{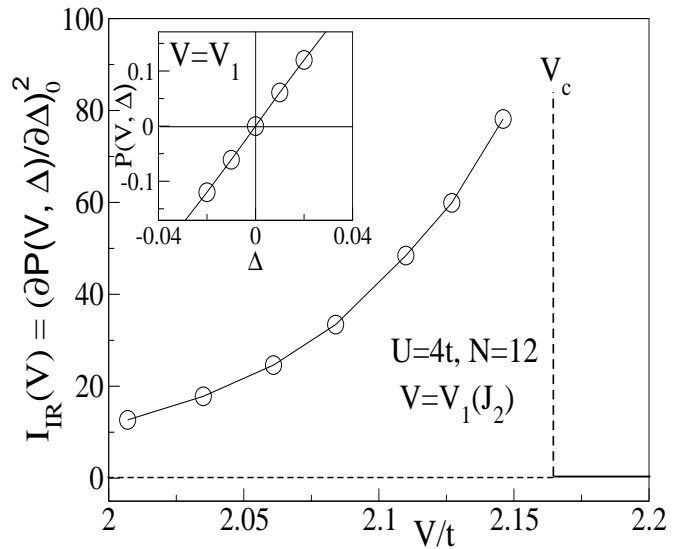}} \\
\caption{  Infrared intensity $I_{IR}(V)$, Eq. \ref{eq13}, of molecular vibrations due to $\Delta$ 
in Eq. \ref{eq12} and broken electronic symmetry in the BOW phase of the EHM. The inset is the 
polarization $P(\Delta)$ in Eq. \ref{eq9}. $I_{IR}$ vanishes by symmetry for $V > V_c$ or $< V_s$.}
\label{fig5}
\end {center}
\end{figure}

We compute the derivative in Eq. \ref{eq13} in the BOW phase at $V = V_1(J_2)$. 
The imaginary part of $Z_N$ is initially proportional to $\Delta$, 
as shown in the inset of Fig. \ref{fig5}. $I_{IR}(V)$ increases rapidly with $V$ 
up to $V_c$, as expected for decreasing $E_J$, and vanishes abruptly in the 
CDW phase where $C_i$ symmetry is restored and e-h symmetry is broken. 
The corresponding band result \cite{r33,r32} for 
$(\partial P(\delta,\Delta)/\partial \Delta)^2_0$ 
increases as $1/\delta^2$ and diverges at the metallic point $\Delta = \delta = 0$ 
where, however, $P$ is not defined. Since the real part of $Z_N(V) = 0$ at $V = V_c(N)$, 
the finite system also has divergent $(\partial P(V,\Delta)/\partial \Delta)_0$ at $V_c(N)$ 
and the BOW phase again resembles a band with small $\delta$, with two major differences. 
First, $I_{IR}(V)$ increases with $V$ up to $V_c$ while $B(V)$ is almost constant. 
Second, $\delta_{eff} $ for IR intensity {\it decreases} with increasing $V$ up to 
$V_c$ while $\delta_{eff}$ for $B(V)$ {\it increases} with $V$ from $V_s$. The SSH model 
has a single band gap of $4\delta t$ instead of the EHM's multiple threshold excitations 
in Table \ref{tb1} and Fig. \ref{fig1}. For example, the metallic point $V_c$ has 
large $E_m \approx 0.5t$.\\

\begin{figure}[h]
\begin {center}
\hspace*{-0cm}{\includegraphics[width=8.5cm,height=9.5cm,angle=-90]{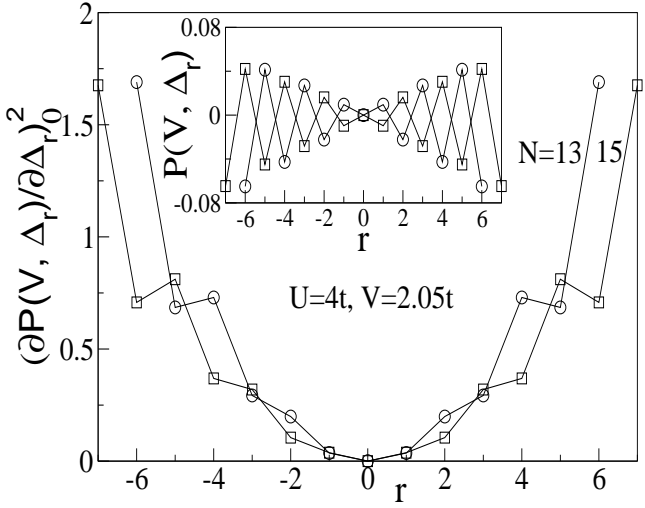}} \\
\caption{ Infrared intensity $I_{IR}(V)$, Eq. \ref{eq13}, of molecular vibrations due 
to $\Delta$ in Eq. \ref{eq12} in 13 and 15-site spin solitons with $V = 2.05t$ and 
$t_{1N} = 0$ in 
Eq. \ref{eq1}. The inset shows $P(V,\Delta_r)$ for $\Delta_r$ = 0.30 in Eq. \ref{eq9} at 
site $r$, with $r = 0$ at the center and $r=\pm (N-1)/2$ at the ends.} 
\label{fig6}
\end {center}
\end{figure}

Domain walls introduce inversion centers that reduce $I_{IR}$ between regions with 
opposite $B$. More quantitatively, we again consider systems with odd $N$, $V = 2.05$ 
and $t_{1N} = 0$. We break $C_i$ symmetry with $\pm \Delta$ at sites $\pm r$ from the 
center and evaluate $ (\partial P/\partial \Delta_r)_0$. Since $\Delta$ in Eq. \ref{eq12} 
applies all sites in Fig. \ref{fig5}, we scale $P(V,\Delta_r)$ by $N/2$ for the soliton. 
The inset to Fig. \ref{fig6} shows $P(V,\Delta_r)$ at $V = 2.05t$ for $\Delta_r = 0.05$, 
with $P = 0$ at $r = 0$ and scaled $P \approx 0.06$ at the ends of a spin soliton. 
The same pattern is found for charge solitons (data not shown) with about $50 \%$ higher 
intensity. In either case, $I_{IR}(V)$ at the ends is an order of magnitude less than 
the $V_1(J_2)$ values for degenerate gs in Fig. \ref{fig5}. 
This finite-size effect can be traced to higher $E_J$ in the radical. 
We did not solve $N = 17$ since finite $\Delta$ doubles the dimensions 
of the many-electron basis. IR intensities are consistent with 
a soliton and an inversion center between regions with $ \pm B(V)$. \\

We consider next the Peierls instability. The SSH model is the generic 
case for e-ph coupling $\alpha = (dt/du)_0$ in a harmonic 1D lattice 
with force constant $k$. Linear coupling is retained in generalizations \cite{r18} 
that require numerical methods and include electron correlation, quantum 
fluctuations, nonadiabatic or 3D effects, spin chains and models with site energies. 
Quantum fluctuations are particularly important for small $\delta $ that is easily 
reversed locally. Fluctuations reduce but do not wash out $\delta \approx 0.18$ 
in the SSH model of polyacetylene \cite{r35}. \\

Now Eq. \ref{eq1} reads

\begin{eqnarray}
H(\delta)  &=& H_{el} +\sum^N_{p=1,\sigma} \delta_p t(  a^{\dagger}_{p,\sigma}  a_{p+1,\sigma} + h.c) \nonum \\
&+& \sum_p \delta^2_p/2\epsilon_d.
\label{eq14}
\end{eqnarray}
with $\epsilon_d = \alpha^2/k$  and constant $\delta_p = (-1)^p \alpha u/k$ in the gs. The distribution {$\delta_p$} is subject to the constraint $\sum_p\delta_p = 0$ for a fixed chain length. The gs energy per site is

\begin{eqnarray}
\epsilon_T(\delta)=\epsilon_0(\delta)+\delta^2/2\epsilon_d.
\label{eq15}
\end{eqnarray}
A minimum at $ \delta \ne 0$ implies a dimerized gs. A {\it global}  adiabatic 
approximation has $\delta(T) > 0$ for $T < T_P$, the Peierls temperature, while a  {\it local } adiabatic 
approximation leads to domains walls between regions of opposite 
$\delta$ for $T > 0$. \\

Our discussion is limited to $\epsilon_T(\delta)$ in Eq. \ref{eq15}. 
The gs of correlated models is unconditionally dimerized when the 
regular array has $E_m = 0$ and $\chi_d(\delta) = -(\partial^2 \epsilon_0/\partial \delta^2)$ 
diverges at $\delta = 0$. Examples \cite{r26} include the ionic phase of organic 
charge transfer salts and the spin fluid phase of Hubbard models or 
Heisenberg spin chains. Dimerization is {\it conditional } at $\epsilon_d \chi_d(0) = 1$ 
when $\chi_d(0)$ is finite in the neutral phase of CT salts or 
the CDW phase of the EHM.\\ 

\begin{figure}[h]
\begin {center}
\hspace*{-0cm}{\includegraphics[width=8.5cm,height=10.5cm,angle=0]{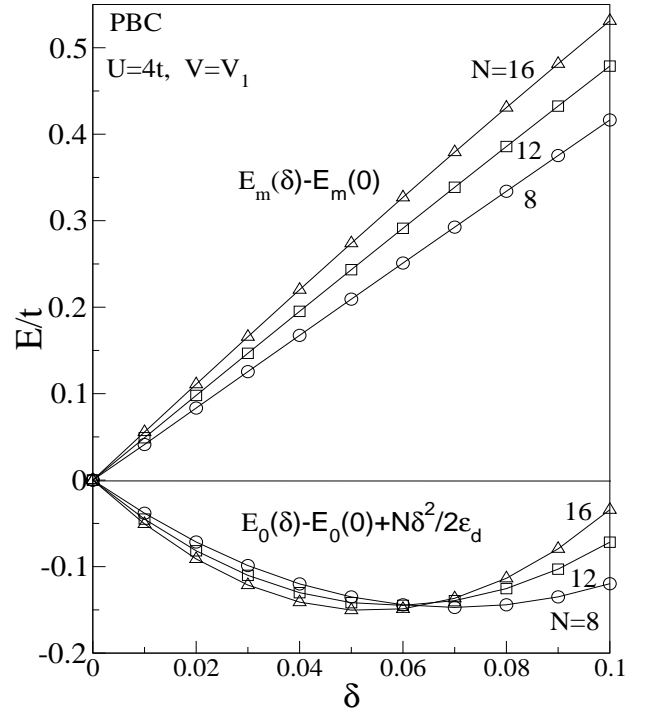}} \\
\caption{ Ground state energy $E_0$ in Eq. \ref{eq15} and magnetic gap $E_m$ of an EHM with dimerization $\delta$ 
in Eq. \ref{eq14} and $\epsilon_d = 0.12$. }
\label{fig7}
\end {center}
\end{figure}

Degenerate gs is different, as seen for $N\epsilon_T(\delta)$ 
in Fig. \ref{fig7} for $V = V_1(N)$ in Table \ref{tb1} and inverse stiffness $\epsilon_d = 0.12$. 
Now $\epsilon_0(\delta)$ goes as  $-B|\delta| - \chi_d(0)\delta^2/2$. The cusp 
$B(V)$ is due to degeneracy at $\delta = 0$ while the energy gap $E_3$ in the singlet sector, 
listed in Table \ref{tb2}, ensures finite $\chi_d(0)$. Minimization of $\epsilon_T$ yields 

\begin{eqnarray}
\delta_{eq}=\pm \epsilon_d B(V)/(1-\epsilon_d \chi_d(0))
\label{eq16}
\end{eqnarray}
with $\delta_{eq} > 0$ for $\delta > 0$ and $\delta_{eq} < 0$ for $\delta < 0$. 
Dimerization is unconditional and increases $E_m(\delta)$ beyond $E_m(0)$ 
as shown in Fig. \ref{fig7} for a vertical excitation. The gs cusp appears 
again because the lowest triplet is not degenerate and hence evolves as $\delta^2$. 
The slope of $(\partial E_m/ \partial \delta)_0$ at the origin is $NB(V_1,N)$. It follows 

that two domain walls that change the $ \delta_p$ pattern without 
changing $\sum_p \delta_p$ lead to lower $E_m$, and reduced $E_m$ are 
readily found for $N = 12$ or 16. Generalization of $2W_S(N)$ in Eq. \ref{eq6} 
is more suitable for spin solitons between regions with opposite $\delta_{eq}$. 
Such simulations are beyond the scope of the present study.\\

An adiabatic approximation gives a dimerized gs with $\pm \delta_{eq}$ 
that has to be relaxed locally for solitons. X-ray detection 
requires that $\delta_{eq}$ not be too small compared to the zero point amplitude, $\langle \delta^2 \rangle^{ 1/2}$, of the Peierls 
phonon $\hbar \omega_P$ in Eq. \ref{eq15},

\begin{eqnarray}
\langle \delta^2 \rangle =\epsilon_d \hbar \omega_P/2t 
\label{eq17}
\end{eqnarray}
Typical values \cite{r36} of $ \hbar \omega_P \approx 100 cm^{-1}$, $\epsilon_d \approx 0.30$ 
and $t \approx 1500 cm^{-1}$ 
in organic stacks return $\langle \delta^2 \rangle^{1/2}  \approx 0.10$ at $T = 0$ 
that increases with $T$. Finite $E_m$ and small $\delta_{eq}$ minimize contributions 
from thermal excitations, in contrast to SSH or correlated Peierls systems in 
which small $\delta_{eq}$ necessarily implies small $E_m$ that vanishes at $\delta = 0$. 
An arbitrarily small perturbation selects one of the degenerate gs of a BOW phase at $0$ K, 
but there is no long-range order in 1D at finite $T$. Solitons lower the free energy and 
result in local adiabatic approximations. At low $T$, the mean separation $R(T)$ between solitons 
is large compared to their widths $2\xi$. Each soliton can then be centered on any of $R(T)$ 
sites. The soliton density in an extended system with $R > 2\xi$ and $N \rightarrow \infty$ is

\begin{eqnarray}
\rho(T)=\frac{R(T)}{N}=exp(\frac{-W}{k_BT})
\label{eq18}
\end{eqnarray}
where $k_B$ is the Boltzmann constant and $W = W_S$ for spin solitons or $W_C$ for charge solitons. 
The $N = 17$ results in Fig. \ref{fig4} are not sufficient for estimating $2\xi$, 
but indicate $2\xi > 30$ for spin solitons and $2\xi \approx 30$ for charge solitons, 
consistent with the SSH estimate \cite{r11} of $2\xi \approx 15$ for larger $\delta = 0.18$. 
Since $W_S < W_C$, spin solitons are thermally accessible and the condition $R(T) > 2\xi$ holds up to $\rho(T) \approx 1\%$.  
\section{Spin susceptibility} 

\begin{figure}[h]
\begin {center}
\hspace*{-0cm}{\includegraphics[width=8.5cm,height=9.5cm,angle=-90]{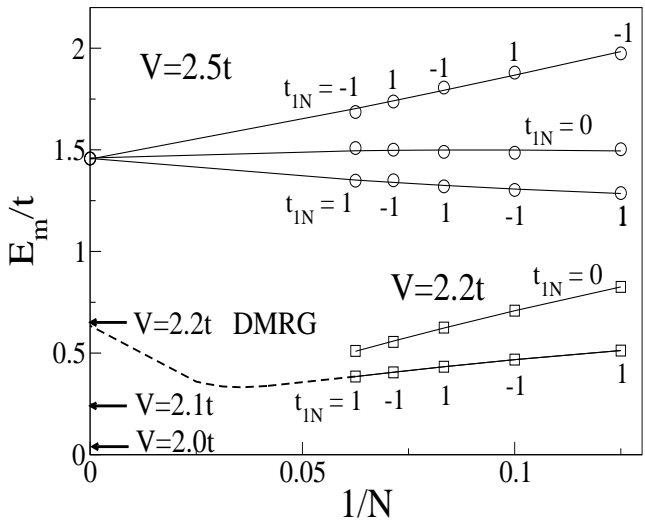}} \\
\caption{ Magnetic gap $E_m$ of the $N$-site EHM, Eq. \ref{eq1}, with $U = 4t$ and the indicated 
V's. Boundary conditions with $t_{1N} = 0$ and $\pm 1$ are discussed in the text. 
The DMRG energy gaps are from ref. \cite{r10}, with minimum $E_m$ around $N \approx 30$ for $V = 2.2t$. }
\label{fig8}
\end {center}
\end{figure}

The magnetic gap $E_m$ opens at the boundary $V_s$ of the spin-fluid and BOW phases, 
and it does so very slowly at a Kosterlitz-Thouless transition \cite{r1}. DMRG with PBC 
provides an independent calculation \cite{r10} showing that $E_m(V)$ opens at $V_s = 1.86t$ for the 
EHM with $U = 4t$. The gap is only $0.023t$ at $V = 2.0$ before reaching $0.241t$ 
at $V = 2.10$ and $0.63t$ at $V = 2.20$, just in the CDW phase. We are 
interested in large $E_m$ near the boundary $V_c$ of the BOW and CDW phases 
in Fig.\ref{fig1}. As noted above, $V_c$ is a metallic point with 
bond orders $p(V)$ in Eq. \ref{eq2} close to $2/\pi$, the band limit. 
The band limit returns $E_m = 0$ for $N = 4n$, when the degenerate orbitals 
at $\epsilon = 0$ are half filled, and $E_m = 4tsin(\pi/N)$ for $N = 4n + 2$. 
Open boundary conditions with $t_{1N} = 0$ have intermediate $E_m$. 
The same pattern is seen in Fig. \ref{fig8} for the EHM at $V = 2.5$, with smallest $E_m(N)$ 
for $N = 4n$, $t_{1N} = 1$ and $N = 4n + 2$, $t_{1N} = - 1$. 
Quite unusually, $E_m$ increases with $N$ at $V = 2.5$. DMRG calculations \cite{r10} of $E_m$ 
with PBC have minimum $E_m(N)$ at $N \approx 30$ for $V = 2.2$. Exact $E_m(N)$ in 
Fig. \ref{fig8} with $t_{1N} = 0$, $\pm 1$ and $V = 2.2$ decrease with $N$. 
Rapidly increasing $E_m$ at $V_c$ is also seen \cite{r10} for other potentials and 
has important implications for modeling the spin susceptibility. In particular, 
larger $N$ is not automatically better.\\

The molar magnetic susceptibility $\chi_M(T)$ allows quantitative comparisons \cite{r13,r24p,r37}
for organic ion-radical solids with small spin-orbit coupling and $g$-factors 
close to the free-electron value, $g = 2.00236$. We take $\chi_M(T)$ 
to be the spin susceptibility after standard corrections for diamagnetism and 
impurities. The full spectrum of Eq. \ref{eq1} is now required, and charge degrees of 
freedom vastly increase the number of states. We extend exact results for $\chi_M(T)$ 
up to $N = 10$. The partition function of the EHM in Eq. \ref{eq1} with even $N$ is
\begin{eqnarray}
Q_N(T)=\sum^{N/2}_{S=0} \sum_{r=1} (2S+1) exp(-E_{Sr}(N)/k_BT) 
\label{eq19}
\end{eqnarray}
The singlet gs is the zero of energy, $E_{01}(N) = 0$, and $E_{Sr}(N)$ 
refers to the state $r$ with spin $S$. The molar spin  susceptibility is 
\begin{eqnarray}
\chi_M(T,N)&=&\frac{N_Ag^2\mu^2_B}{3tN Q_N} \big(\frac{t}{k_BT} \big)\sum^{N/2}_{S=0} \sum_{r=1}S(S+1) \nonum 
\\ & & \times (2S+1) exp(-E_{Sr}(N)/k_BT) 
\label{eq20}
\end{eqnarray}
where $\mu_B$ is the Bohr magneton and $N_A$ is Avogadro's number. 
Finite $E_m = E_{11}$ leads to $\chi_M(0) = 0$, in contrast to finite 
$\chi_M(0)$ for $V < V_s$ in the spin-fluid phase. We set $E_{\sigma} = 0$ 
in the BOW phase and consider finite $N$ to be a coarse-grained 
approximation of an extended system with a dense spectrum for $E \ge E_m$.\\

\begin{figure}[h]
\begin {center}
\hspace*{-0cm}{\includegraphics[width=8.5cm,height=9.5cm,angle=-90]{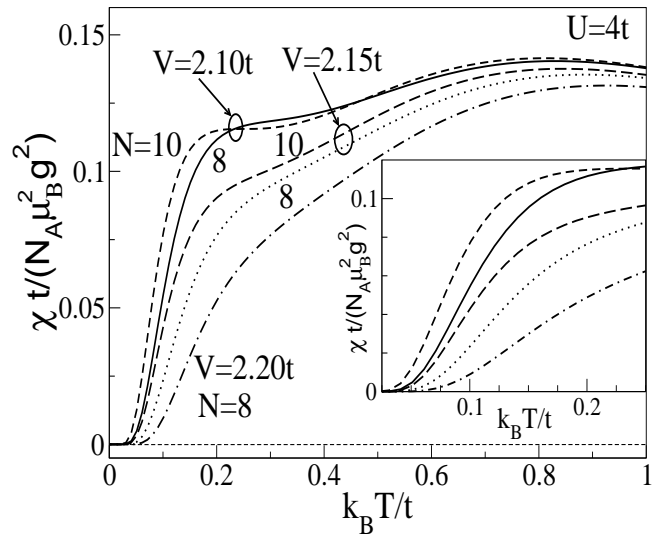}} \\
\caption{Temperature dependence of the molar spin 
susceptibility $\chi_M(T)$ of $N$-site EHM with $U = 4t$ 
and $V$ in Eq. \ref{eq1} The curves in the inset are for the same $U$, $V$. }
\label{fig9}
\end {center}
\end{figure}

Figure \ref{fig9} shows $\chi_M(T)$ up to $k_BT = t$ for $N = 8$ with PBC ($t_{1N} = 1)$ and $N = 10$ with $t_{1N} = -1$ 
at $V = 2.10$, 2.15 and 2.20. Finite-size effects become negligible at high $T$, as has long been recognized in 
spin chains. In addition, the broad plateau around $k_BT \approx t$ depends weakly on $V$,
 which simply reflects the narrowness of the BOW phase. By contrast, the $T \approx 0$ behavior is 
governed by $E_m(V)$ and is very sensitive to $V$ at constant $U$, $t$. Although $t \approx 10^3 K$ 
is a small electronic energy, experiment is limited to much lower $T$. The inset of Fig. \ref{fig9} 
expands the relevant range. Finite $E_m$ suppresses $\chi_M$ at low $T$ after 
which $\chi_M$ is almost linear in $T$. The BOW phase of a frustrated spin chain \cite{r38}  without charge degrees 
of freedom also has an almost linear $\chi_M(T)$ and a broad maximum. \\

A BOW phase with $E_m > 0$ has $\chi_M(T) \approx 0$ up to $T \approx T_1$ 
followed by a linear increase for $T > T_1$. For example, Rb-TCNQ(II) has \cite{r19} $T_1 \approx 140$ K 
that roughly fixes $t(V)$ in the inset to Fig. \ref{fig9} and completely specifies $\chi_M(T)$ 
of the EHM with $U = 4t$. Although $\chi_M(T)$ increases almost linearly to $T = 300 K$, 
the magnitude at 300 K rules out the $V > 2.10t$ curves in the inset while the $V = 2.10$ 
curve with $k_BT/t \approx 0.15$ at 300 K fails for $T_1$. An EHM with $U = 4t$ is 
not quantitative for Rb-TCNQ(II). Improved fits are possible for small 
$\delta_{eq} \approx 0.01$ but such modeling also entails 
variation of Coulomb interactions.\\

The EHM boundary $V_s$ between the spin fluid and BOW 
phases has been difficult to model and has been reported \cite{r5} 
to be as high as $V_s \approx 2.02t$ at $U = 4t$. Since $\chi_M(0)$ is finite 
in the spin fluid phase and $E_m$ opens slowly for $V > V_s$, very different $\chi_M(T)$ 
are then calculated in the BOW phase \cite{r5}. The evolution of $E_m(V,N)$ in Fig. \ref{fig8} 
with $N$ and DMRG results support the original estimate \cite{r1} of $V_s \approx 1.86t$ at $U = 4t$ 
based on excited-state crossovers in Table \ref{tb1}. Rapidly increasing $E_m(V)$ as $V$ 
approaches the metallic point $V_c$ is found in the BOW phase of related Hubbard models \cite{r10}. 
Large $E_m$ is needed for the Rb-TCNQ(II) susceptibility as well as for K and 
Na-TCNQ at high $T$ where X-ray structures \cite{r44} indicate regular $ {\rm TCNQ^{-} }$ stacks.\\ 

\section {Discussion}
Rice \cite{r27} recognized the possibility of measuring e-mv coupling constants $g_n$ 
from polarized IR spectra when $C_i$ symmetry is broken on dimerization. Several 
groups \cite{r28,r29} extended the procedure to extracting transferable $g_n$ for selected $\pi$-donors and 
$\pi$-acceptors. K-TCNQ was a prime example \cite{r40} of a crystal with dimerized ${ \rm TCNQ^-}$ stacks 
at 300 K. Polarized mid-IR spectra show coupled ts modes that are shifted to the 
red from the corresponding Raman transitions \cite{r40,r28}. Powder Rb-TCNQ(II) has virtually 
identical IR transitions \cite{r41} whose polarization along the stack has been 
confirmed in single crystals \cite{r42}, but strikingly different temperature 
dependence. Raman spectra and $g_n$ of ${\rm TCNQ^- }$ are expected 
to be the same, since solid-state perturbations are usually small \cite{r28,r29}.\\

The 100 and 295 K crystal structures of Rb-TCNQ(II) decisively indicate \cite{r19} 
a regular stack of $\rm TCNQ^-$ at inversion centers and interplanar separation $R = 3.174 \AA$. 
The 295 K structure is in excellent agreement with previous data \cite{r39p}, 
and the triclinic space group $\rm {\bf \it P\bar{1}}$ is retained at $100 K$. Low $R$ factors 
and examination of thermal ellipsoids at 100 K place a conservative limit on 
dimerization of $R_+ - R_- < 0.05 \AA$ \cite{r19}. Yet negligibly small $\chi_M(T)$ below 140 K 
implies a large $E_m$ and IR data indicates broken electronic $C_i$ symmetry. Broken $C_i$ 
symmetry in a BOW phase accounts naturally for large $E_m$ near the CDW boundary and for IR intensity 
at 0 K. We take up the different temperature dependence of the Rb salt.\\

The IR intensity in Eq. \ref{eq13} goes as $(\partial P(V,\Delta)/\partial \Delta)^2_0$ \cite{r20,r26} 
and increases with $V$ in the BOW phase as shown in Fig. \ref{fig5}. The intensities 
for a spin soliton in Fig. \ref{fig6} are smaller and vanish at the center. Each 
soliton introduces a $C_i$ center between regions with $\pm B(V)$ and reduces 
$(\partial P(V,\Delta)/\partial \Delta)_0$ over $\approx 2 \xi$ sites. We approximate the 
temperature dependence as

\begin{eqnarray}
\frac{I_{IR}(T)}{I_{IR}(0)}=\frac{1}{(1+2\xi \rho_S(T))}.
\label{eq21}
\end{eqnarray}

\begin{figure}[h]
\begin {center}
\hspace*{-0cm}{\includegraphics[width=8.5cm,height=8.5cm,angle=-90]{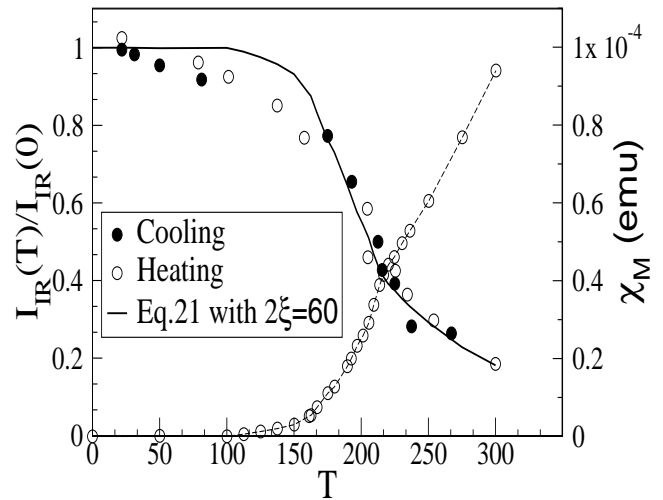}} \\
\caption{Intensity ratio $I_{IR}(T)/I_{IR}(0)$ of the 722 ${\rm cm^{-1}}$ mode of 
Rb-TCNQ(II) on cooling (open symbols) and heating (closed symbols) from ref \cite{r43}. 
The solid line is Eq. \ref{eq21} with $2\xi = 60$ and spin density $\rho_S(T)$ from $\chi_M(T)$ in ref. \cite{r19}}.
\label{fig10}
\end {center}
\end{figure}

\noindent Here $\rho_S(T)$ is the spin density given by $\chi_M(T)/\chi_C$, where  
$\chi_C = N_Ag^2\mu^2_B/4k_BT$ is the Curie susceptibility. 
The intensity is 50\% lower at $2\xi \rho_S(T) = 1$ where spin solitons 
overlap. Eq. \ref{eq21} relates two measured quantities through 
$2\xi $, as shown in Fig. \ref{fig10} for $2\xi = 60$, which is in the 
expected range. The $\chi_M(T)$ data is for Rb-TCNQ(II) 
from ref. \cite{r19} and gives $\rho_S(T)$. The intensity ratio of the 
722 $\rm  cm^{-1}$ mode is from Fig. \ref{fig2} of ref. \cite{r43}, with open and 
closed symbols on cooling and heating. Similar $T$ dependence is seen for other 
mid-IR modes of crystals \cite{r42}. The fit supports a BOW phase interpretation and 
electronic solitons rather than a specific microscopic model or parameters. 
A microscopic model must account for $\chi_M(T)$ in addition to the intensity ratio.\\

We have examined the BOW phase of the EHM at intermediate correlation $U = 4t$ by 
direct solution of Eq. \ref{eq1} for finite $N$. We used degenerate gs at $V = V_1(N)$ in Table \ref{tb1} 
to break inversion symmetry in finite systems and varied $J_2$ in Eq. \ref{eq5} 
to scan $V_1(N)$ over the BOW phase. Exact degeneracy enforced by $J_2$ turns out 
to be important for e-mv coupling to Holstein phonons but not for the order parameter 
$B(V) = p_+ - p_-$ in Eq. \ref{eq3}. We compared BOW properties due 
to electronic correlation in a regular 1D chain to the SSH model of a dimerized
 band. SSH results for topological solitons carry over for many aspects of spin and 
charge solitons in the BOW phase, albeit with different $\delta_{eff}$ for different properties. 
We focused on the consequences of a degenerate gs and finite $E_m$, especially large $E_m$ 
close to the metallic point $V_c$. Broadly similar results are expected in BOW phases of 
other Hubbard-type models with intermediate correlation.\\ 

In the adiabatic approximation for the lattice, linear e-ph coupling 
generates a dimerized gs in both the BOW phase of the EHM and the SSH model, 
but they are different. The SSH model has a standard Peierls transition at $T_P$. 
Thermal population of excited states stabilizes the regular array for $T > T_P$, 
and low $T_P$ necessarily implies weak coupling or a stiff lattice. 
The Peierls transition of polyacetylene is far above its thermal stability, and we 
are not aware of evidence for $\delta_{eq}(T)$ variations up to $\approx 400$ K. 
Spin-Peierls systems illustrate decreasing $\delta_{eq}(T)$ up to $T_{SP} < 20 $ K 
that can be modeled in the adiabatic approximation \cite{r39}. The BOW phase of the 
EHM at intermediate $U = 4t$ samples a different sector of parameter space, 
one in which substantial $E_m$ up to $ \approx 0.5t$ suppresses thermal excitations. 
Spin solitons in Eq. \ref{eq18} with $W_S(\delta)$  
give a small $\chi_M(T)$ in this range, and $E_m = 2W_S$ remains finite at $\delta = 0$. 
The BOW phase has novel aspects that need further study. The principal theoretical issues are 
quantum fluctuations or nonadiabatic phonons that may suppress a sharp Peierls transition when e-ph
coupling is weak. The experimental problem is to detect small dimerization against a background 
of zero-point motions.\\

The present discussion is limited to the BOW phase of the EHM at $U = 4t$. 
We have developed the consequences of coupling to lattice phonons and to 
molecular vibrations in the adiabatic approximation. Similar results are expected \cite{r10} 
for other quantum cell models with electron-hole symmetry and will be needed to model physical 
systems with BOW phases, starting with Rb-TCNQ(II). Since alkali-TCNQ salts are semiconductors, 
they have Coulomb interactions rather than a Hubbard $U$ and are stabilized close to the CDW boundary 
by the 3D electrostatic (Madelung) energy \cite{r24}. The Na and K-TCNQ salts have dimerization phase 
transitions \cite{r44} with some 3D character since the cations also dimerize. 
The regular structure at high $T$ has small $\chi_M(T)$ that increases with $T$ in a manner that suggests a BOW phase. 
Both $\pi-$radical organic stacks and conjugated polymers are quasi-1D systems whose 
initial modeling is without interchain interactions. \\

In summary, we have characterized the BOW phase of the EHM with intermediate $U = 4t$ 
by exact treatment of finite systems with degenerate gs at $V = V_1(N)$. The 
elementary excitations are electronic solitons, both spin and charge, in a regular array. 
Solitons in the correlated BOW phase resemble the familiar solitons of the 
SSH model for e-ph coupling in a tight-binding band. Several measures indicate 
an ``effective'' dimerization $\delta < 0.10$ that is considerably less 
than $ \delta = 0.18$ for the SSH model of polyacetylene. Charge fluctuations 
are coupled to molecular (Holstein) phonons that become IR active in 
the BOW phase due to broken $C_i$ symmetry. The $T$ dependence of 
IR modes is consistent with spin solitons whose width is $2\xi \approx 60$ lattice constants. 
The BOW phase is dimerized at 0 K in the adiabatic approximation, 
but gs degeneracy and finite $E_m$ lead to novel aspects for a possible Peierls 
transition. Previous discussions of SSH solitons greatly facilitate analysis 
of the BOW phase. So have previous treatments of e-ph and e-mv coupling in 1D Hubbard 
models for conjugated polymers and organic ion-radical or charge-transfer crystals. 
The BOW phase of Hubbard-type models with intermediate correlation has some unique aspects 
that invite further study as well as features that are common to such models. \\

{ \bf Acknowledgments:} ZGS thanks A. Girlando for access to unpublished 
IR spectra and A. Painelli for discussions of polarization and BOW phases. 
We thank the National Science Foundation for partial support of this work 
through the Princeton MRSEC (DMR-0819860).

\end{document}